\documentclass[
 reprint,
superscriptaddress,
 amsmath,amssymb,
 aps,
]{revtex4-2}

\usepackage{graphicx}
\usepackage{dcolumn}
\usepackage{bm}
\usepackage{xcolor} 
\usepackage{upgreek}
\usepackage{hyperref}
\hypersetup{
    colorlinks=true,
    citecolor=blue,
    linkcolor=blue,
    filecolor=blue,      
    urlcolor=blue,
    pdftitle={Overleaf Example},
    pdfpagemode=FullScreen,
    }
\begin{document}
\preprint{APS/123-QED}
\title{Optical coherence storage in dark exciton states using spin-dependent three-pulse photon echo}

\author{I.~A.~Solovev}
\author{R.~S.~Nazarov}
\affiliation{St. Petersburg State University, 198504 St. Petersburg, Russia}
\author{A.~A.~Butiugina}
\affiliation{St. Petersburg State University, 198504 St. Petersburg, Russia}
\author{S.~A.~Eliseev}
\author{V.~A.~Lovcjus}
\author{Yu.~P.~Efimov}
\affiliation{St. Petersburg State University, 198504 St. Petersburg, Russia}
\author{Yu.~V.~Kapitonov}
\affiliation{St. Petersburg State University, 198504 St. Petersburg, Russia}
\author{I.~A.~Yugova}\email{i.yugova@spbu.ru}
\affiliation{Spin Optics Laboratory, St. Petersburg State University, 198504 St. Petersburg, Russia}
\date{\today}

\begin{abstract}
Extension of coherent response time is a desired goal in the field of all-optical information processing implemented in classical and quantum ways. Here we demonstrate how spin-dependent stimulated photon echo can be used to extend decay time of coherent signal from exciton ensemble. We experimentally studied photon echoes from excitons in a model single InGaAs/GaAs quantum well subject to transverse magnetic field. Field-induced quantum beats lead to oscillation of exciton population between bright and long-lived dark excitons. As a result, photon echo decays much longer with decay time attaining dark exciton lifetime in case of non-oscillatory regime.

\end{abstract}

\maketitle

\section{Introduction}
Coherence time is a key feature in the field of information processing implemented either in all-optical classical or in quantum way [1]. Therefore the desired goal is to enhance preservation of coherence. In this context, studies of dark excitons those coherence time should be basically longer due to restriction of radiative recombination are of interest.
To detect the dynamics of dark excitons it was proposed to use a transverse magnetic field in 2D semiconductors MoSe2 [2] and WSe2 [3] which resulted in a significant increase in the radiative lifetime by several orders of magnitude. Various types of illumination of dark excitons have also been studied in quantum wells [4], dots [5], and carbon nanotubes [6]. However, these works did not consider coherent interaction. A protocol for coherent dark exciton spin precession driven by polarized light has been demonstrated in single quantum dots [7]. Manipulation of the quantum state of a dark exciton via a biexciton state without the application of a magnetic field was shown in [8] also in a single quantum dot.
For studying a coherence in exciton ensembles, the photon echo technique is attractive because it allows one to overcome the reversible phase relaxation of the system associated with the inhomogeneous broadening of any ensemble. In its basic form, two-pulse spontaneous photon echo is the simplest protocol for implementing optical memory [9,10]. In addition, photon echo can be used as a tool for studying the coherent optical properties of semiconductor nanostructures: measuring the phase relaxation time [11], determining the energy structure of the systems under study [12]. Another protocol for storing optical information is three-pulse stimulated photon echo, where the polarization created by the first pulse is translated into population by the second pulse, and then back into polarization by the third one. As a result, the signal provides information about the population decay time, $T_1$. In the absence of additional external influences $T_1$ is a characteristic time for storing optical information while maintaining coherence. In the absence of non-radiative relaxation, the time $T_1$ is equal to the radiative recombination time $\tau_r$. The efficiency of interaction with light is inversely proportional to the time $\tau_r$, and therefore the contradiction between an energy efficiency, a speed and a storage duration of optical information arises. A possible way to resolve this contradiction is to transfer the information recorded in the material optical excitation into a spin subsystem that does not interact with the optical field. In [13], a protocol was proposed for recording optical information into the spin subsystem in an experiment on stimulated photon echo with the application of an external transverse magnetic field in n-doped CdTe/CdMgTe quantum wells. As a result of the experiment, it was possible to increase the duration of optical memory by more than three orders of magnitude from picoseconds to tens of nanoseconds. Despite the impressive results, the main disadvantage of the described method is the need to apply fairly high magnetic fields (on the order of 1 T), which makes it necessary to use an external superconducting magnet. In this work, we propose a new optical memory protocol that allows, in a stimulated photon echo experiment, to transfer information to the spin subsystem in a low magnetic field. In this case, the proposed approach is applied not to a system of trions, the detection of which requires a special structure design and a special excitation mode, but to a system of neutral excitons.

\section{Theory}
\label{sec:theory}
\begin{figure}
    \centering
    \includegraphics[width=\linewidth]{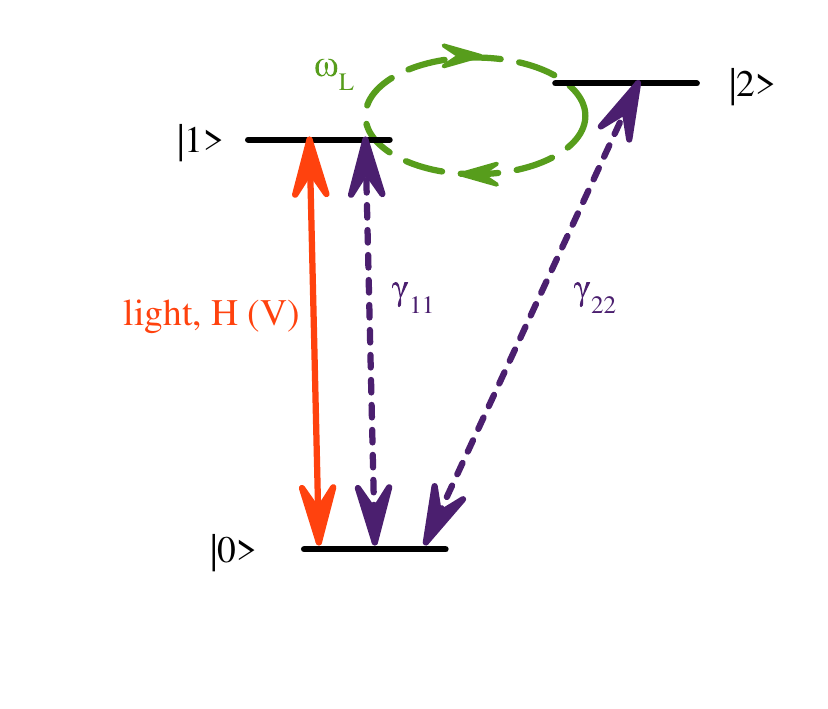}
    \caption{Schematic representation of the energy levels of the system under consideration. Optical transition between states $\left|0\right\rangle - \left|1\right\rangle$ is allowed. States $\left|1\right\rangle$ 
        and $\left|2\right\rangle$ are related by precession with frequency $\omega_L$. Relaxation of the populations of states $\left|1\right\rangle$ and $\left|2\right\rangle$ to the ground state occurs at rates $\gamma_{11}$ and $\gamma_{22}$, respectively.}
    \label{fig:scheme}
\end{figure}

To analyze the three-pulse photon echo protocol for an ensemble of excitons, a three-level model system consisting of the ground state $\left|0\right\rangle$ and excited states $\left|1\right\rangle$ and $\left|2\right\rangle$ coupled by precession with the frequency $\omega_L$ is used(see Fig.~\ref{fig:scheme}). Only the transition $\left|0\right\rangle - \left|1\right\rangle$ is optically allowed. For excitons in A3B5 and A2B6 quantum wells this situation is realized, for example, with parallel or perpendicular linear polarizations of the laser pulses to the direction of the magnetic field lying in the plane of the sample [14, 15]. In this case, the states $\left|1\right\rangle$ and $\left|2\right\rangle$ are split by the exchange interaction value $\delta_0$. And the Larmor precession frequency is defined either as the sum of the electron and hole precession frequencies $\omega_L=\omega_e+\omega_h$, if the polarization of light is parallel to the direction of the magnetic field, or as the difference $\omega_L=\omega_e-\omega_h$, if the polarization is perpendicular to the direction of the magnetic field.

This problem was considered in the density matrix formalism. We assumed that the duration of light pulses is much shorter than the characteristic relaxation times and precession periods of the system under study, which is in good agreement with experimental conditions in most cases [9-15].
Under such assumptions, we can divide the problem of solving the Lindblad equation for the density matrix $i\hbar\frac{\partial}{\partial t}\rho(t)=\left[H_0+V(t),\rho\right]+i\hbar\mathrm{\Gamma}$ into two parts: we consider separately the interaction with light, without taking into account precession in the magnetic field and relaxation during the action of the pulse and the dynamics of the system in a magnetic field.

Basic Hamiltonian of the system in a magnetic field (Vocht geometry):
\begin{equation}  
H_0=\left(\begin{matrix}0&0&0\\0&\hbar\omega_0+\frac{\delta_0}{2}&\frac{\hbar\omega_L}{2}\\0&\frac{\hbar\omega_L}{2}&\hbar\omega_0-\frac{\delta_0}{2}\\\end{matrix}\right) 
\end{equation}
Here $\hbar\omega_0$ is the energy of the allowed optical transition, $\delta_0$ is the value of the exchange interaction in the exciton, $\omega_L$ is the Larmor precession frequency.

Hamiltonian of the interaction with light:
\begin{equation}  
V(t)=\frac{\hbar}{2}\left(\begin{matrix}0&f^\ast e^{i\omega t}&0\\fe^{-i\omega t}&0&0\\0&0&0\\\end{matrix}\right) 
\end{equation}

Here $f(t)$ is the pulse envelope: $f(t)=-\frac{{2e}^{i\omega t}}{\hbar}\int{d\left(\bf{r}\right)E\left(\bf{r},t\right)d^3\bf{r}}$, $d\left(\bf{r}\right)$ is the dipole moment of the optical transition, $\omega$ is the frequency of the exciting light. For simplicity, we consider resonant excitation ($\omega_0=\omega$) and a rectangular pulse with duration $t_p$.

We write the relaxation as:
\begin{equation}
\Gamma=\left(\begin{matrix}\rho_{11}\gamma_{11}+\rho_{22}\gamma_{22}&-\rho_{01}\gamma_1&-\rho_{02}\gamma_2\\-\rho_{10}\gamma_1&-\rho_{11}\gamma_{11}&-\rho_{12}\gamma_{12}\\-\rho_{20}\gamma_2&-\rho_{21}\gamma_{21}&-\rho_{22}\gamma_{22}\\\end{matrix}\right)
\end{equation}
Here $\rho_{ik}$ are the elements of the density matrix in the basis
 $\left\{\left|0\right\rangle, \left|1\right\rangle, \left|2\right\rangle\right\}$, 
$\left\{i,k=0,1,2\right\}$. 
$\gamma_1$ and $\gamma_2$ are relaxation rates of the coherences $\left|0\right\rangle - \left|1\right\rangle$ and |$\left|0\right\rangle - \left|2\right\rangle$. $\gamma_{11}$ and $\gamma_{22}$ are population relaxation rates, and $\gamma_{12}=\gamma_{21}=(\gamma_{11}+\gamma_{22})/2$ are relaxation rates of their mutual coherence. For simplicity, we will not consider processes that can make additional contributions to $\gamma_{12}$.

In experiments measuring three-pulse photon echo, information about the optical pulse is transferred to the population and spin coherence of the system. Indeed, the analysis shows that for the system under consideration, the signal amplitude is determined by the dynamics of the difference between the populations of state $\left|1\right\rangle$ and state $\left|0\right\rangle$ and the coherence of states $\left|1\right\rangle$ and $\left|2\right\rangle$ during the time between the second and third pulses $t_{23}$:
\begin{eqnarray}
P_{SPE}(t_{12},t_{23})&\sim & K_1(t_{12})\left(\rho_{11}\left(t_{23}\right)-\rho_{00}\left(t_{23}\right)\right) \\ \nonumber
&+& K_2(t_{12})\rho_{12}(t_{23}) \nonumber
\end{eqnarray}
Coefficients $K_1$ and $K_2$ depend on the excitation parameters, system parameters and dynamics of coherences $\left|0\right\rangle - \left|1\right\rangle$ and $\left|0\right\rangle - \left|2\right\rangle$ during the time between the first and the second pulses arrivals, $t_{12}$.

\begin{figure*}[hbt] 
\begin{center}
  \includegraphics[width=0.48\linewidth]{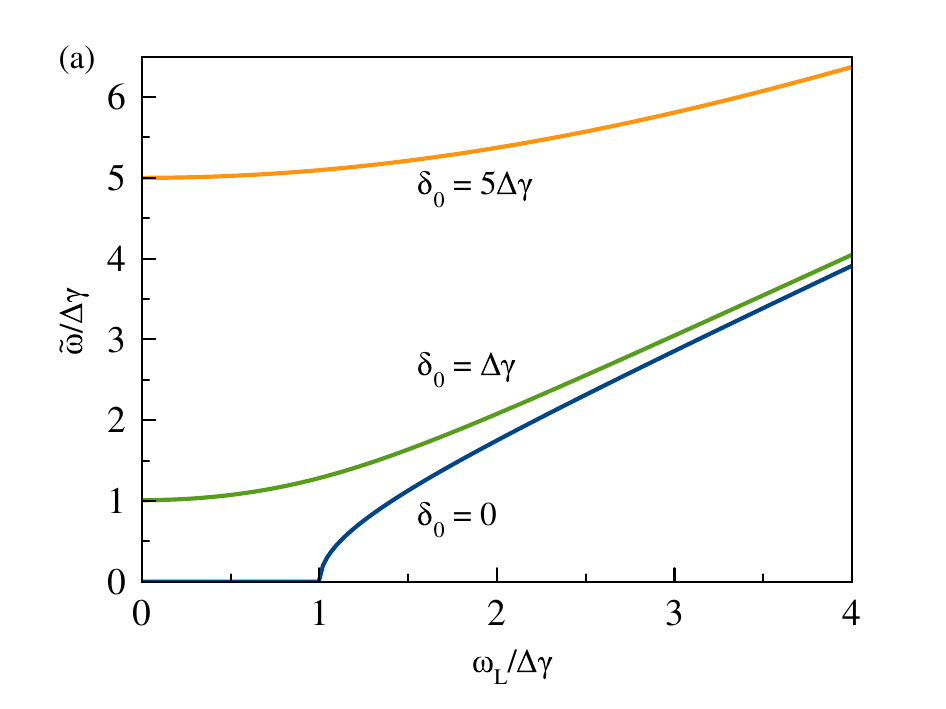}
  \vspace{2em}
  \includegraphics[width=0.48\linewidth]{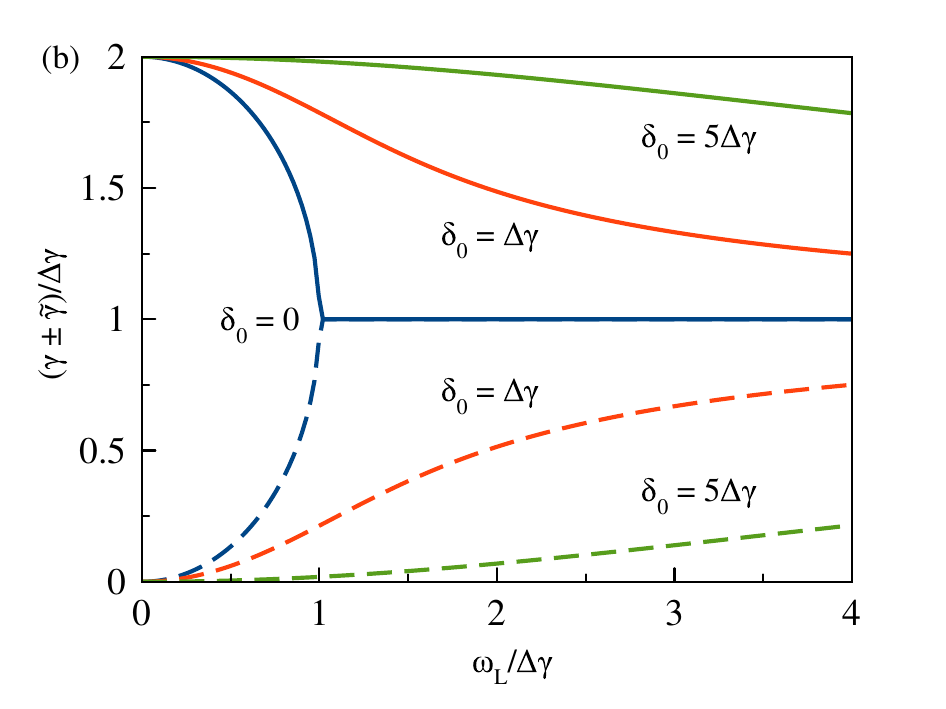}
  \end{center}
  \caption{Dependences of $\widetilde{\omega}$ (a) and $\gamma\pm\widetilde{\gamma}$ (b) on the magnetic field expressed through the precession frequency in dimensionless units ($\omega_L/\Delta\gamma$) for different values of $\delta_0$. For simplicity, $\gamma_{22}=0$. In panel (b), dotted lines indicate the values of $\gamma+\widetilde{\gamma}$, solid lines indicate $\gamma-\widetilde{\gamma}$.}
  \label{fig:omega_gamma}
\end{figure*} 

\begin{figure*}[hbt] 
\begin{center}
  \includegraphics[width=0.48\linewidth]{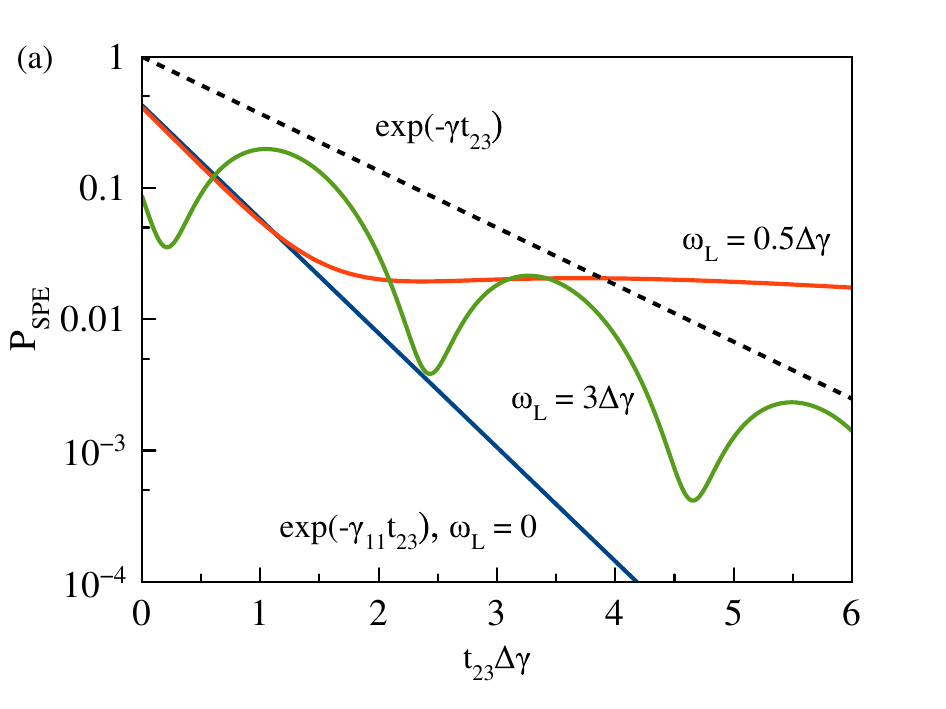}
  \vspace{2em}
  \includegraphics[width=0.48\linewidth]{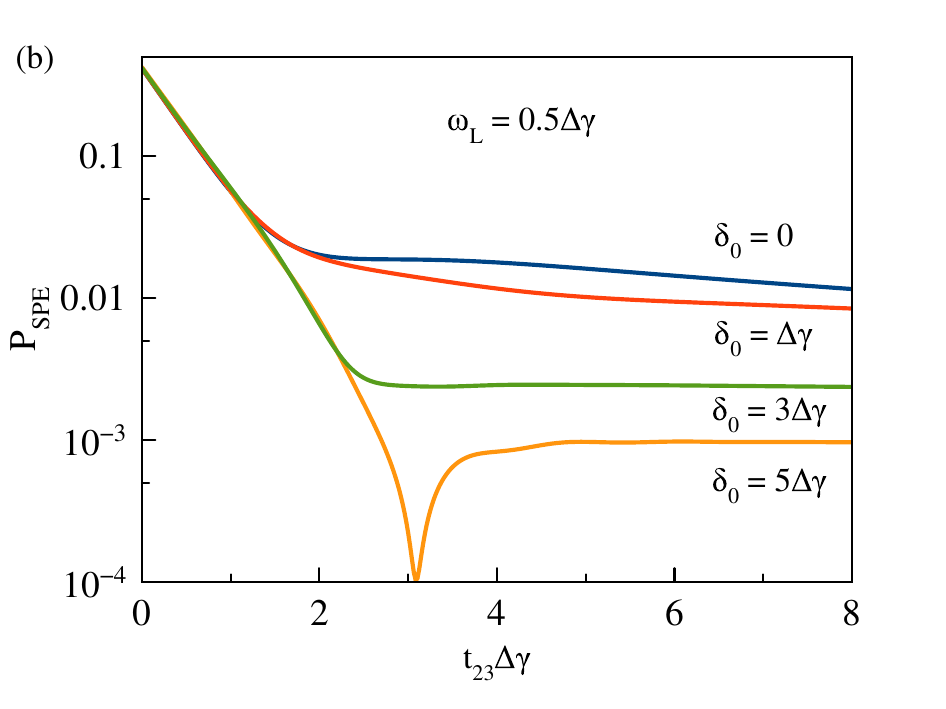}
\caption{(a) Dependence of the photon echo amplitude on the delay between the second and third pulse at $\delta_0=0$. For $\omega_L=0$, a single-exponential decay is observed with the relaxation rate of the population of the state $\left|1\right\rangle$, $\gamma_{11}$ (lower straight line). As $\omega_L$ increases, there is first a transition to an aperiodic regime with a long-lived component (red curve, $\omega_L=0.5\Delta\gamma$), then to oscillations decaying at a rate of $\gamma$ (green curve, $\omega_L=3\Delta\gamma$). Calculation parameters: $\gamma_{22}=0, \gamma_1=0.5\gamma_{11}, \gamma_2=0.1\gamma_{11}, t_{12}=0.7/\gamma_1$. The values of $\gamma_1$, $\gamma_2$ and $t_{12}$ are taken from the analysis and fitting of experimental data, see the next paragraph. (b) Dependence of the amplitude of the photon echo on the delay between the second and third pulse $t_{23}$ at $\omega_L=0.5\Delta\gamma$ for different values of $\delta_0$.}  
  \end{center}
  \label{fig:tdynamics}
  \end{figure*}

In zero magnetic field the system under consideration is two-level system. After excitation by light the population of the system $\rho_{11}$  relaxes at a characteristic rate $\gamma_{11}$ (Fig.~3(a), blue curve, $\omega_L=0$). From general considerations it is clear that in a magnetic field the dynamics of the system becomes more complicated as a result of the admixture of an optically inactive (“dark”) state. The total lifetime of the system in the excited state should be extended and oscillations are possible at a frequency determined by the Larmor frequency precession $\omega_L$ and exchange interaction $\delta_0$.
A more detailed analysis shows that different relaxation rates $\gamma_{11}$ and $\gamma_{22}$ lead to renormalization of precession frequency and decay rates in a magnetic field. 

In this case, the renormalized precession frequency $\widetilde{\omega}=Re(\Omega)$, possible decay rates: $\gamma$, $\gamma \pm Im(\Omega)\equiv \gamma \pm \widetilde{\gamma}$, where  $\Omega\equiv\sqrt{\left(\omega_L\right)^2+\left(\frac{\delta_0}{\hbar}-i\Delta\gamma\right)^2}$, $\Delta\gamma\equiv(\gamma_{11}-\gamma_{22})/2$, $\gamma\equiv(\gamma_{11}+\gamma_{22})/2$.

For the amplitude of the three-pulse photon echo signal, an analytical expression can be obtained:
\begin{widetext}
\begin{equation} 
P_{SPE}\sim K_{123}\left[A_2\left(A_3\left(-2{{|M}_1|}^2-\left|L\right|^2\right)\left|R_1\right|^2-iM_1L^\ast R_1R_{12}^\ast\right)+iA_3R_1^\ast R_{12}\left(2M_1^\ast L-M_2L^\ast\right)-{|L|}^2\left|R_{12}\right|^2\right],
\label{eq:SPE1}
\end{equation}
$K_{123}=\frac{if_1{f^\ast}_2{f^\ast}_3}{\left|f_1\right|\left|f_2\right|\left|f_3\right|}A_1B_1B_2B_3e^{-(\gamma_1+\gamma_{2\ })t_{12}}e^{-\gamma\ t_{23}}\rho_{00}\left(0\right)$, 
\end{widetext}
$A_{1,2,3}=\cos\left(|f_{1,2,3}|t_p\right)$, $B_{1,2,3}\equiv \sin\left(|f_{1,2,3}|t_p\right)$, the indices indicate the pulse number. $t_p$ is pulse duration.

\begin{eqnarray}
M_1&\equiv &\cos{\left(\frac{\Omega t_{23}}{2}\right)}+\alpha\sin{\left(\frac{\Omega t_{23}}{2}\right)},\nonumber \\ \nonumber
M_2&\equiv &\cos{\left(\frac{\Omega t_{23}}{2}\right)-}\alpha\sin{\left(\frac{\Omega t_{23}}{2}\right)},\\ \nonumber
L&\equiv &\beta\sin{\left(\frac{\Omega t_{23}}{2}\right)}, \\ \nonumber
\alpha &\equiv &-i\left(\frac{\delta_0}{\hbar}-i\Delta\gamma\right)/\Omega, \beta\equiv\frac{\omega_L}{\Omega}.\nonumber
\end{eqnarray}

The values of $R_1$ and $R_{12}$ are determined by the dynamics of the polarization of the system between the first two pulses (see Appendix~\ref{Appendix}).

This expression can be rewritten more compactly by grouping the exponents:
\begin{eqnarray}
P_{SPE}&\sim & Q_1\left(t_{12}\right)e^{-\left(\gamma+\widetilde{\gamma}\right)t_{23}}+Q_2\left(t_{12}\right)e^{-\left(\gamma-\widetilde{\gamma}\right)t_{23}} \\ \nonumber
&+& Q_3\left(t_{12},t_{23}\right)e^{-\gamma t_{23}},
\label{eq:SPE}
\end{eqnarray}
Here $Q_3$ is a function of $cos\left(\widetilde{\omega}t_{23}\right)$ and $sin\left(\widetilde{\omega}t_{23}\right)$.
 
It is especially clear to demonstrate the contributions of these three components, which decay at different rates if $\delta_0$ is neglected. In this case, the polarization dynamics are divided into two very different regimes. Fig.~\ref{fig:omega_gamma}(a,b) shows the dependences of $\widetilde{\omega}$ and $\gamma\pm\widetilde{\gamma}$ on the magnetic field expressed through the precession frequency in dimensionless units ($\omega_L/\Delta\gamma$) for different values of $\delta_0$. It is clear from the figures that at $\delta_0=0$ in magnetic fields $\omega_L<<\ \Delta\gamma$ there cannot be a precession of the photon echo amplitude ($\widetilde{\omega}=0$), but there is only decay with three characteristic rates: $\gamma$ and $\gamma\pm\widetilde{\gamma}$.

For $\omega_L\ll\Delta\gamma$ and $\delta_0=0$: 
\begin{equation}
\gamma+\widetilde{\gamma}\approx\gamma-\Delta\gamma\left(1-\frac{\omega_L^2}{2{\Delta\gamma}^2}\right)=\gamma_{22}+\frac{\omega_L^2}{2\Delta\gamma},
\nonumber
\end{equation}
\begin{equation}
\gamma-\widetilde{\gamma}\approx\gamma+\Delta\gamma\left(1-\frac{\omega_L^2}{2{\Delta\gamma}^2}\right)=\gamma_{11}-\frac{\omega_L^2}{2\Delta\gamma}.
\end{equation} 

Thus, in a low magnetic field, instead of oscillations in the photon echo signal, an aperiodic regime with long decay can be observed, the time of which is determined by the lifetime of the dark state $\frac{1}{\gamma+\widetilde{\gamma}}\approx\frac{1}{\gamma_{22}}$ (see Fig.3(a), red curve, $\omega_L=0.5\Delta\gamma$).

In high magnetic fields, when $\omega_L>\{\Delta\gamma, \gamma\}$, $\widetilde{\gamma}=0$ (Fig.~\ref{fig:omega_gamma}(b)). In this case, the photon echo signal decays at a rate of $\gamma$ (Fig.3(a)).

From Fig.~\ref{fig:omega_gamma}(a,b) it is clear that at $\delta_0\neq 0$ and for low magnetic fields $\widetilde{\omega}$ differs from zero. Therefore, the oscillating component will contribute to the amplitude of the photon echo. However, its decay rate is $\gamma$ (Eq.~\ref{eq:SPE}) and the long-lived contribution will still be determined by the rate of the slowest process: $\gamma+\widetilde{\gamma}$. In addition, from Fig.~\ref{fig:omega_gamma}(b) it follows that an increase in $\delta_0$ expands the range of $\omega_L$ values for which the $\gamma+\widetilde{\gamma}\approx\gamma_{22}$ approximation works well. That means $\delta_0$ manifests itself as an effective longitudinal magnetic field that stabilizes long-lived dynamics. Fig.~3(b) shows the dependence of the amplitude of the photon echo on the delay $t_{23}$ at $\omega_L=0.5\Delta\gamma$ for different values of $\delta_0$. The shift in the signal inflection and the preservation of the long-lived nature of the decay are clearly visible.
  
 Expression~(\ref{eq:SPE1}) allows us to simulate the behavior of the spin-dependent stimulated photon echo signal from an exciton system when excited by linearly polarized light at an arbitrary value of the magnetic field and exchange interaction constant. 
 
\section{Experimental details} 
\begin{figure}
    \centering
    \includegraphics[width=\linewidth]{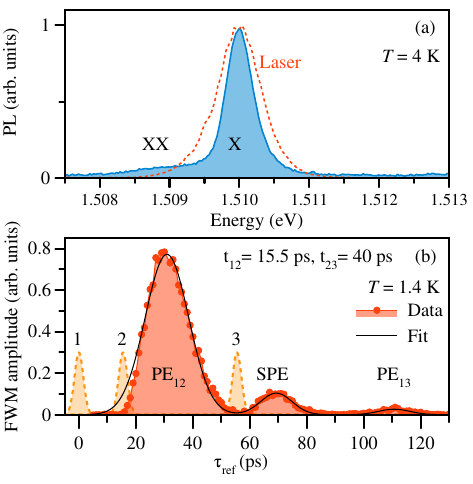}
    \caption{(a) Photoluminescence spectrum of the quantum well under study. (b) Temporal signals of four-wave mixing. The peaks labeled PE$_{12}$ and PE$_{13}$ correspond to the two-pulse photon echo from pulses 1-2 and 1-3, respectively. SPE corresponds to a three-pulse photon echo signal.}
    \label{fig:pl_and_pe}
\end{figure}

To test the theoretical predictions and to demonstrate different decay regimes of the three-pulse photon echo signal, a model system was chosen, namely a high-quality 3 nm InGaAs/GaAs quantum well with a 3\% In concentration in the well layer. Excitons with a heavy hole in such structures, firstly, have a small inhomogeneous broadening comparable to the uniform linewidth (radiative width $\hbar \Gamma_R$ of the spectral line is 35 $\mu$eV, non-radiative broadening $\hbar \Gamma_{NR}$=70 $\mu$eV), see Fig.~\ref{fig:pl_and_pe}(a). Secondly, the coherence time of light excitons reaches 30 ps. The $T_2$ time of dark excitons for the structure under study was detected using the spin-dependent two-pulse photon echo protocol and amounted to 250 ps (for more details see [15]).

The sample was characterized using photoluminescence spectroscopy at a temperature of $T = 4-10$~K, in order to do this the sample was placed in a closed-cycle helium cryostat. For optical excitation, a laser with a photon energy tuned to the exciton resonance in the bulk layer was used, focused on the sample into a spot $50x50$ micrometers in size.
The main research method was the measurement of degenerate four-wave mixing. The use of a Tsunami tunable titanium-sapphire picosecond laser as an optical excitation source made it possible to obtain a spectral resolution of 0.5 meV, while maintaining sufficient time resolution of a few picoseconds. The sample was excited by a sequence of picosecond laser pulses and the time delay between pulses was set by optical delay lines. The signal was mixed using a beam splitter cube with an additional reference pulse and the interference term was detected by means of a balanced photodetector. Scanning the delay of the reference pulse allowed to obtain the cross-correlation signal between the four-wave mixing and the reference pulse with picosecond resolution. Optical heterodyning was used to increase the signal-to-noise ratio many times (for more details, see Ref.~[15]).
Laser pulses following the second and third channels were mixed using a non-polarizing splitter and focused on the sample at the same angle ($k_2=k_3$), which made it possible to measure a two-pulse photon echo and a three-pulse photon echo under the same conditions. The measurements were carried out using linear excitation and detection polarizations. The polarization configuration of the experiment is further denoted in the text as $ABC\rightarrow D$, where $A$, $B$, $C$ are the polarizations of the exciting pulses, and $D$ is the detected polarization of the signal. The main polarizations of the experiment: horizontal (H), parallel to the direction of the magnetic field $B$ (the Voigt geometry), and vertical (V), orthogonal to the magnetic field. To study spin-dependent effects in the photon echo signal, superconducting magnets were used, which made it possible to achieve a transverse magnetic field in the Voigt geometry of up to 5 Tesla.

\section{Experimental results and discussion}

The photoluminescence spectrum shown in Fig.~\ref{fig:pl_and_pe}(a) has two pronounced peaks corresponding to a neutral exciton with a heavy hole $X$ and a biexciton $XX$. By precisely tuning the narrow-spectrum pulse to the resonance of the exciton with a heavy hole we can study the exciton only and exclude biexciton states from consideration.

\begin{figure*}[hbt]
    \centering
    \includegraphics[width=\linewidth]{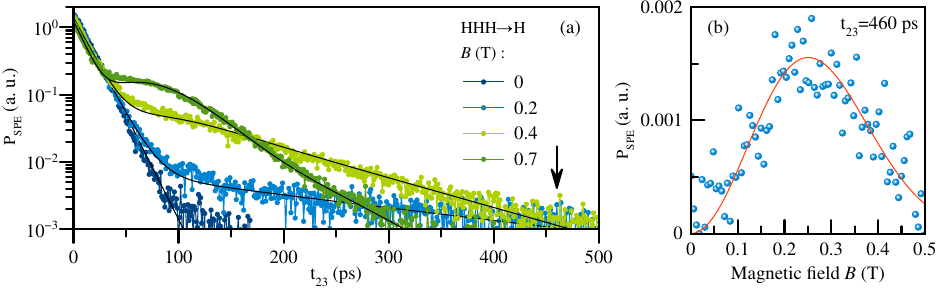}
    \caption{(a) Dependence of the three-pulses photon echo signal on the delay time between the second and the third pulses for different values of the external magnetic field. Noisy curves are an experiment, black smooth curves are the result of modeling using formula~(\ref{eq:SPE1}). (a) Dependence of the three-pulses photon echo signal on the magnetic field at the delay indicated by the black arrow on the left panel, $t_{23} = 460$~ps. The points are experimental data, the curve is the result of modeling using Eg.~(\ref{eq:SPE1}).}
    \label{fig:fit1}
\end{figure*}

Figure~\ref{fig:pl_and_pe}(b) shows the profiles of four-wave mixing signal during excitation to X-resonance with delays between pulses $t_{12}$=15.5~ps and $t_{23}$=40 ps. The temporal profile of the four-wave mixing signal in the $2k_2-k_1$ direction consists of three Gaussian peaks, two of them correspond to the two-pulses photon echo (PE12: $t_{ref}=31$~ps, PE13: $t_{ref}=$111 ps) and the third is the three-pulses photon echo (SPE: $t_{ref}= 69$~ps).
One can notice the asymmetry of the profile of the first photon echo and a slight deviation of the maximum of the three-pulses photon echo from that expected in the case of a model of a two-level system excited by delta pulses [16]. The time resolution of the setup allows us to detect the signal at the time $t_{ref}=2t_{12}+t_{23}$, therefore we can filter the three-pulses photon echo from the two-pulses photon echo and track the decay kinetics when scanning the delay $t_{23}$. The signal decays monoexponentially with a characteristic population decay time $T_1= 15$~ps (see Fig.~\ref{fig:fit1}(a)).

One can see from Fig.~\ref{fig:fit1}, if a transverse magnetic field is applied then the signal kinetics changes significantly: the population decay time increases and an oscillatory character appears. The measurements clearly demonstrate the manifestation of coherent dynamics of dark excitons, namely the mixing of the populations of bright and dark exciton levels due to the Larmor precession of the exciton spin in the transverse field as predicted by theory (see Sec.~\ref{sec:theory}). There is good agreement between the model and experimental data. It can be seen that with increasing magnetic field, an aperiodic regime of oscillations of the three-pulses photon echo is observed. Figure~\ref{fig:fit1}(b) shows the behavior of the amplitude of a long-lived echo signal with increasing magnetic field strength. The increase and subsequent decrease are due to the dependence of the decay time of the long-lived component on the field.

\begin{figure}
    \centering
    \includegraphics[width=\linewidth]{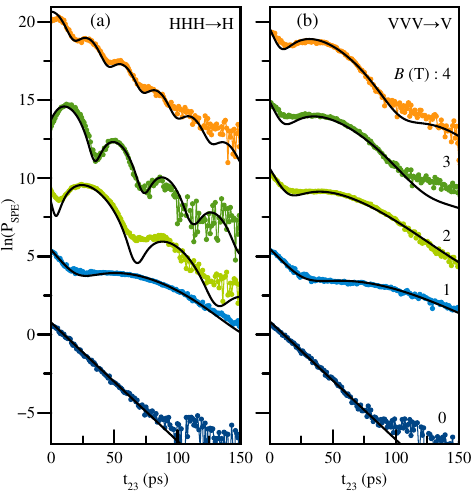}
    \caption{Dependences of the amplitude of the three-pulses photon echo signal on the delay between the second and third pulses for different magnetic field values indicated for each curve. The panels show the dependences for polarization configurations when the orientations of linear polarizations of pulses and photon echoes coincide with the direction of the magnetic field ($HHH\rightarrow H$) or are perpendicular to it ($VVV\rightarrow V$).}
    \label{fig:fit2}
\end{figure}

Figure~\ref{fig:fit2} shows the time dependence of the echo signal amplitude on the magnetic field over a larger range of fields in two different polarization configurations. As can be seen from the right panel of Fig.~\ref{fig:fit2} the aperiodic regime can be created not only by applying a small magnetic field but also by choosing the polarization of the optical excitation. In the case when the polarizations of pulses and of the echo are orthogonal to the direction of the external magnetic field the oscillation frequency can remain low in a large field due to partial compensation of the precession of the electron and hole spins. Measurements of the decay kinetics of a two-pulses photon echo in several polarimetric configurations make it possible to determine the electron and hole $g$ factors in the plane of the structure with good accuracy. Herewith information about $g$ factors is used in the analysis of experimental data to isolate various components in the signal in aperiodic regime. Kinetics measured in high fields make it possible to separate the contributions of the inhomogeneous spread of g factors and the exchange interaction between the electron and hole in the exciton to the echo signal.

A detailed analysis of the dependences of the photon echo on the polarization configuration and the magnitude of the magnetic field using expression~(\ref{eq:SPE1}) allows us to extract the in-plane electron $g$ factor $|g_{e,\bot}| = 0.44 \pm 0.05$. The obtained value of the $g$ factor coincides with that for thin InGaAs/GaAs quantum wells [17,18].
The value of the hole $g$-factor and its nonlinear dependence on the magnetic field is consistent with our previous studies [15]. The corresponding in-plane heavy hole $g$ factor becomes comparable to the electron one at small magnetic field and reaches a maximum value of $|g_{h,\bot}| \approx 0.3 $. The magnitude of the exchange interaction is also consistent with the data from [15], $\delta_0 \le 10 \mu$~eV. An analysis of the decay kinetics of the photon echo signal in low magnetic fields gives a value for the lifetime of the population of the dark state $T_{22} = 170$~ps, that is an order of magnitude greater than than the lifetime of bright excitons. The obtained value for time $T_{22}$ is in good agreement with theoretical estimates and experimental data [19,20,21]. Thus, the application of a spin-dependent photon echo protocol for an ensemble of excitons can be used to increase the optical coherence time of the system.

\section{Conclusions}
In our work, we used a comprehensive spin-dependent, stimulated photon echo protocol to increase the effective coherence time of an ensemble of excitons in an InGaAs/GaAs quantum well. A model was developed to describe the decay kinetics of the stimulated photon echo amplitude. Analytical expressions for the amplitude of the photon echo signal were obtained. From a theoretical analysis the optimal experimental conditions for the transition of the system to a long-lived dark state were determined. A comparative analysis of the predictions of the developed theory and the obtained experimental data was carried out and good agreement was obtained.
From the analysis of kinetics, the frequencies of Larmor precession of electron and hole spins were extracted. The scatter of g-factors and value of the exchange interaction are estimated, and the degree of influence of these parameters on the shape of the coherent response is determined.
Based on the results of the developed theory, numerical simulation and experimental data, experimental protocols are proposed for optimizing long-lived regimes in the photon echo signal on an ensemble of excitons in low-dimensional semiconductor structures.

\begin{acknowledgments}
This work was supported by Grant of the Russian Science Foundation No 22-22-00439 https://rscf.ru/en/project/22-22-00439/. This work was carried out on the equipment
of the SPbU Resource Center “Nanophotonics”.
\end{acknowledgments}
\section{Appendix: Additional theoretical aspects}
\label{Appendix}
We assume that the duration of light pulses is much shorter than the characteristic relaxation times and precession periods of the system and divide the problem of solving the Lindblad equation for the density matrix into two parts: we consider separately the interaction with light, without taking into account precession in the magnetic field and relaxation during the action of the pulse; and then separately the dynamics of the system in a magnetic field.

Solution for the elements of the density matrix, describing the action of a rectangular light pulse of duration $t_p$ at resonant excitation:
\begin{widetext}
\begin{eqnarray}
\rho_{01}(t_p)&=&e^{i\omega_0t_p}\left[A^2\rho_{01}\left(0\right)+\left(\frac{f^\ast}{\left|f\right|}\right)^2B^2\rho_{10}\left(0\right)+\frac{if^\ast}{\left|f\right|}AB\left(\rho_{00}\left(0\right)-\rho_{11}\left(0\right)\right)\right], \\ \nonumber
\rho_{02}(t_p&=&e^{i\omega_0t_p}\left[A\rho_{02}\left(0\right)-\frac{if^\ast}{\left|f\right|}B\rho_{12}\left(0\right)\right], \\ \nonumber
\rho_{00}(t_p)&=&A^2\rho_{00}\left(0\right)+B^2\rho_{11}\left(0\right)+\frac{i}{\left|f\right|}AB\left({f\rho}_{01}\left(0\right)-{f^\ast\rho}_{10}\left(0\right)\right), \\ \nonumber
\rho_{11}(t_p)&=&B^2\rho_{00}\left(0\right)+A^2\rho_{11}\left(0\right)-\frac{i}{\left|f\right|}AB\left({f\rho}_{01}\left(0\right)-{f^\ast\rho}_{10}\left(0\right)\right), \\ \nonumber
\rho_{12}(t_p)&=&A\rho_{12}\left(0\right)-\frac{if}{\left|f\right|}B\rho_{02}\left(0\right),\\ \nonumber
\rho_{22}(t_p)&=&\rho_{22}(0). \nonumber
\end{eqnarray}
\end{widetext}
Here $A\equiv cos\left(|f|t_p\right)$, $B\equiv sin\left(|f|t_p\right)$, $\rho_{ik}\left(0\right)$ are the initial conditions for the elements of the density matrix.

Dynamics in a magnetic field, solution for the coherence of states $\left|0\right\rangle - \left|1\right\rangle$ and $\left|0\right\rangle - \left|2\right\rangle$:
\begin{widetext}
\begin{eqnarray}
\rho_{01}(t)&=&e^{i\omega_0t}e^{-\frac{\gamma_1+\gamma_2}{2}\ t}\left[R_1^\ast\left(t\right)\rho_{01}\left(0\right)+R_{12}^\ast\left(t\right)\rho_{02}\left(0\right)\right],\\ \nonumber
\rho_{02}(t)&=&e^{i\omega_0t}e^{-\frac{\gamma_1+\gamma_2}{2}\ t}\left[R_{12}^\ast\left(t\right)\rho_{01}\left(0\right)+R_2^\ast\left(t\right)\rho_{02}\left(0\right)\right].\\ \nonumber
\end{eqnarray}
\begin{eqnarray}
R_1\left(t\right)&\equiv &\cos{\left(\Omega_{nd}t/2\right)}-i{(\delta}_0/\hbar-i\Delta\gamma_{nd})\left(\frac{\sin{\left(\Omega_{nd}t/2\right)}}{\Omega_{nd}}\right),\\ \nonumber
R_{12}\left(t\right)&\equiv &-i\omega_L\left(\frac{\sin{\left(\Omega_{nd}t/2\right)}}{\Omega_{nd}}\right)\\ \nonumber
R_2\left(t\right)&\equiv &\cos{\left(\Omega_{nd}t/2\right)}+i{(\delta}_0/\hbar-i\Delta\gamma_{nd})\left(\frac{\sin{\left(\Omega_{nd}t/2\right)}}{\Omega_{nd}}\right) \\ \nonumber
\Omega_{nd}&\equiv &\sqrt{\left(\frac{\delta_0}{\hbar}-i\Delta\gamma_{nd}\right)^2+\left(\omega_L\right)^2}, \Delta\gamma_{nd}\equiv\gamma_1-\gamma_2. \nonumber
\end{eqnarray}
\end{widetext}

For convenience and uniformity with subsequent formulas, here the direct and conjugate quantities $R_i$ and $R_i^\ast$ are introduced in reverse with respect to the Refs.~\onlinecite{Solovev2021,Solovev2022}.

Populations and coherence of states $\left|1\right\rangle$ and $\left|2\right\rangle$ in a magnetic field:
\begin{widetext}
\begin{eqnarray}
\rho_{11}\left(t\right)&=&e^{-\gamma\ t}\left({{|M}_1\left(t\right)|}^2\rho_{11}\left(0\right)+{|L\left(t\right)|}^2\rho_{22}\left(0\right)+iM_1\left(t\right)L^\ast\left(t\right)\rho_{12}\left(0\right)-iM_1^\ast\left(t\right)L\left(t\right)\rho_{21}\left(0\right)\right),\\ \nonumber
\rho_{22}\left(t\right)&=&e^{-\gamma\ t}\left(|{M_2\left(t\right)|}^2\rho_{22}\left(0\right)+\left|L\left(t\right)\right|^2\rho_{11}\left(0\right)-iM_2^\ast\left(t\right)L\left(t\right)\rho_{12}\left(0\right)+iM_2\left(t\right)L^\ast\left(t\right)\rho_{21}\left(0\right)\right),\\ \nonumber
\rho_{12}(t)&=&e^{-\gamma\ t}\left(iM_1\left(t\right)L^\ast\left(t\right)\rho_{11}\left(0\right)-iM_2^\ast\left(t\right)L\left(t\right)\rho_{22}\left(0\right)+M_1\left(t\right)M_2^\ast\left(t\right)\rho_{12}\left(0\right)+{|L\left(t\right)|}^2\rho_{21}\left(0\right)\right),\\ \nonumber
\rho_{00}\left(t\right)&=&\rho_{00}\left(0\right)+\rho_{11}\left(0\right)\left(1-e^{-\gamma\ t}\left({{|M}_1\left(t\right)|}^2+{|L\left(t\right)|}^2\right)\right)+\rho_{22}\left(0\right)\left(1-e^{-\gamma\ t}\left({{|M}_2\left(t\right)|}^2+{|L\left(t\right)|}^2\right)\right)\\ \nonumber
&-&ie^{-\gamma\ t}\left(\left(M_1\left(t\right)L^\ast\left(t\right)-M_2^\ast\left(t\right)L\left(t\right)\right)\rho_{12}\left(0\right)-\left(M_1^\ast\left(t\right)L\left(t\right)-M_2\left(t\right)L^\ast\left(t\right)\right)\rho_{21}\left(0\right)\right)
\end{eqnarray}
\end{widetext}

\begin{eqnarray}
M_1\left(t\right)&\equiv & \cos{\left(\frac{\Omega t}{2}\right)}+\alpha\sin{\left(\frac{\Omega t}{2}\right)},\\ \nonumber
M_2\left(t\right)&\equiv &\cos{\left(\frac{\Omega t}{2}\right)}-\alpha\sin{\left(\frac{\Omega t}{2}\right)},\\ \nonumber
L\left(t\right)&\equiv &\beta\sin{\left(\frac{\Omega t}{2}\right)}, \\ \nonumber
\Omega &\equiv &\sqrt{\left(\omega_L\right)^2+\left(d_0-i\Delta\gamma\right)^2},\\ \nonumber
\alpha &\equiv &-i\left(d_0-i\Delta\gamma\right)/\Omega, \\ \nonumber
\beta &\equiv &\frac{\omega_L}{\Omega}, \gamma\equiv\frac{\gamma_{11}}{2}+\frac{\gamma_{22}}{2},
\Delta\gamma\equiv\frac{\gamma_{11}}{2}-\frac{\gamma_{22}}{2}.\nonumber
\end{eqnarray}

Consistently using these expressions, we can obtain an expression for the amplitude of the three-pulse echo signal Eq.~\ref{eq:SPE}.

\end{document}